\documentclass[11pt]{article}
\usepackage{epsfig}
\newcommand{\dalpha}{{\dot{\alpha}}}
\newcommand{\dbeta}{{\dot{\beta}}}
\newcommand{\dgamma}{{\dot{\gamma}}}
\newcommand{\ddelta}{{\dot{\delta}}}

\newcommand{\ft}[2]{{\textstyle\frac{#1}{#2}}\,}
\newcommand{\gf}{{| \Lambda_{L} \rangle}}
\newcommand{\be}{\begin{equation} }
\newcommand{\ee}{\end{equation} }
\newcommand{\bea}{\begin{eqnarray}}
\newcommand{\eea}{\end{eqnarray}}

\newcommand{\su}{\mbox{su}}

\def\H{{\stackrel{\rightarrow}{H}}}
\def\E{{\cal E}}

\def\tr{{\rm tr}}
\def\Tr{{\rm Tr}}
\def\I_M{{I_{\scriptscriptstyle M\times M}}}

\begin{document}

\thispagestyle{empty}
\rightline{~~~AEI-2002-051} \rightline{~~~KIAS-P02042} \rightline{{\tt hep-th/0207061}}

\vskip 2cm \centerline{ \Large\bf  Superalgebra for M-theory on a pp-wave}

\vskip 2cm

\centerline{Nakwoo Kim$^a$  and Jeong-Hyuck Park$^{b}$} \vskip 10mm \centerline{ \it $^a$ Max-Planck-Institute f\"{u}r Gravitationsphysik,
Albert-Einstein-Institut} \centerline{\it Am M\"{u}hlenberg 1, D-14476 Golm, Germany} \vskip 3mm \centerline{\it $^b$ School of Physics, Korea Institute
for Advanced Study} \centerline{ \it 207-43, Cheongryangri-Dong, Dongdaemun-Gu, Seoul 130-012, Corea } \vskip 1.2cm

\centerline{\tt  (kim@aei-potsdam.mpg.de, jhp@kias.re.kr)}

\vskip 1.7cm

\begin{quote}
We study  the superalgebra of the M-theory on a fully supersymmetric pp-wave. We identify the algebra  as  the special unitary  Lie superalgebra,
$\mbox{su}(2|4\,;2,0)$ or $\mbox{su}(2|4\,;2,4)$, and  analyze its root structure. We  discuss the typical and atypical  representations deriving the
typicality condition explicitly in terms of the energy and other four quantum numbers. We  classify the BPS multiplets   preserving $4,\,8,\,12,\,16$
real supercharges and obtain the corresponding  spectrum. We show that in the BPS multiplet either the lowest energy floor  is a $\mbox{su}(2)$ singlet
or the highest energy floor is a $\mbox{su}(4)$ singlet.
\end{quote}

\newpage
\setcounter{footnote}{0}

\section{Introduction}
Recent advances in string/M-theory~\cite{Berenstein:2002jq} show that the matrix model for the M-theory in the infinite momentum frame
\cite{Banks:1996vh} actually belongs to a one-parameter family of matrix models  with the introduction of  a mass parameter, $\mu$. The new massive
matrix model corresponds to a partonic description of $D0$-branes or alternatively  a discretized supermembrane action \cite{deWit:1988ig} in the
maximally supersymmetric pp-wave background \cite{Kowalski-Glikman:wv,Figueroa-O'Farrill}.  It is the presence of the massive terms which makes the model
more accessible as the mass terms lift up the flat directions completely and the perturbative expansion is possible by powers of $\mu^{-1}$
\cite{Dasgupta:2002hx,Kim:2002if}.

An interesting feature of the pp-wave matrix model  is that the supercharges do not commute with the Hamiltonian because the supersymmetry transformations
are explicitly time dependent. Accordingly the bosons and fermions have different masses and numbers as noted in \cite{Berenstein:2002jq}. More
thorough    understanding requires the complete analysis on the superalgebra itself, including the root structure and the representations. The complete
classification of the Lie superalgebras was first done by Kac \cite{Kac:em,Kac:qb}, from which looking at the bosonic symmetry,
$\mbox{su}(2)\oplus\mbox{su}(4)$, one can easily conclude that the \textit{complexified} superalgebra of the pp-wave matrix model is $\mbox{A}(1|3)$.

 In this paper, we elaborate on the method of Kac and analyze  the superalgebra of the M-theory on a fully supersymmetric pp-wave to
demonstrate the root structure explicitly and discuss the typical and atypical  representations. In particular we show that the actual superalgebra of
the pp-wave matrix model is the  \textit{special unitary  Lie superalgebra}, $\mbox{su}(2|4\,;2,0)$ or $\mbox{su}(2|4\,;2,4)$ depending on the sign of
$\mu$, and derive the typicality condition explicitly in terms of the energy and other four quantum numbers. We  also completely classify the BPS
multiplets as $4/16,\,8/16,\,12/16,\,16/16$ and obtain the corresponding  spectrum.

The organization of the paper is as follows. In section \ref{superalgebra}, after setting up the gamma matrices and other conventions,  we write down the
superalgebra of the M-theory on a  pp-wave and identify the algebra  as  the special unitary Lie superalgebra, $\mbox{su}(2|4\,;2,0)$ or
$\mbox{su}(2|4\,;2,4)$. In section \ref{superalgebra}, we  analyze the root structure of the algebra. We also present the quadratic  super-Casimir
operator. Section \ref{IRR} discusses various types of irreducible representations. After describing the  general properties, we discuss the typical and
atypical irreducible representations, and we derive explicitly  the typicality condition  in terms of the energy and other four
$\mbox{su}(2)\oplus\mbox{su}(4)$ quantum numbers. We completely classify the BPS multiplets as a special type of the atypical unitary representation. We
find there are
 BPS multiplets preserving $4,\,8,\,12,\,16$ real supercharges, and obtain the corresponding exact spectrum. Further we show
that in the BPS multiplet either the ``lowest energy''  floor
 is a $\mbox{su}(2)$ singlet or the ``highest energy''  floor  is a $\mbox{su}(4)$ singlet. We list explicit examples in section \ref{example}.
 We conclude in Section \ref{conclusion}.

\textit{Note Added}: Up on finishing this paper, a related work \cite{Keshav} has appeared on the archive but differs in  details.

\section{Superalgebra\label{superalgebra}}
\subsection{Gamma Matrices and Spinors}
To make the $\mbox{SO}(3)\times\mbox{SO}(6)$ structure of the M-theory on a pp-wave manifest,  we write the nine dimensional gamma matrices  in terms of
the three and six dimensional ones, $\sigma^{i},\gamma^{a}$
\begin{equation}
\begin{array}{ll}
\Gamma^{i}=\sigma^{i}\otimes\gamma^{(7)}&~~~\mbox{for~}~i=1,2,3\\
{}&{}\\
\Gamma^{a}=1\otimes\gamma^{a}&~~~\mbox{for~}~a=4,5,6,7,8,9\,.
\end{array}
\end{equation}
With the choice
\begin{equation}
\gamma^{(7)}=i\gamma^{4}\gamma^{5}\cdots\gamma^{9}=\left(\begin{array}{cc}1&0\\0&-1\end{array}\right)\,, \label{gamma7}
\end{equation}
the six dimensional gamma matrices  are in the block diagonal form
\begin{equation}
\begin{array}{cc}
\gamma^{a}=\left(\begin{array}{cc}0&\rho^{a}\\ \bar{\rho}^{a}&0\end{array}\right)\,,~~~&~~\rho^{a}\bar{\rho}^{b}+\rho^{b}\bar{\rho}^{a}=2\delta^{ab}\,.
\end{array}
\end{equation}
Note that the fact $\bar{\rho}^{a}=(\rho^{a})^{\dagger}$   ensures  $\gamma^{a}$ to be hermitian. Further we set the $4\times 4$ matrices, $\rho^{a}$ to
be anti-symmetric~\cite{Park:1998nr}
\begin{equation}
\begin{array}{cc} (\rho^{a})_{\dalpha\dbeta}=-(\rho^{a})_{\dbeta\dalpha}\,,~~&~~(\bar{\rho}^{a})^{\dalpha\dbeta}=-(\bar{\rho}^{a})^{\dbeta\dalpha}\,.
\end{array}
\label{anti-sym}
\end{equation}
Henceforth $\alpha,\beta=1,2$ are the  ${\mbox{su}(2)}$ indices, while $\dalpha,\dbeta$ denote the ${\mbox{su}(4)}$ indices, $1,2,3,4$.

$\{\rho^{ab}\equiv\frac{1}{2}(\rho^{a}\bar{\rho}^{b}-\rho^{b}\bar{\rho}^{a})\}$  forms an orthonormal basis of general $4\times 4$ traceless matrices
\begin{equation}
\tr(\rho^{ab}\rho^{cd})=4(\delta^{ad}\delta^{bc}-\delta^{ac}\delta^{bd}) \label{6dorthonormal}
\end{equation}
having the completeness relation
\begin{equation}
\textstyle{-\frac{1}{8}}(\rho^{ab})_{\dalpha}{}^{\dbeta}(\rho_{ab})_{\dgamma}{}^{\ddelta}+
\textstyle{\frac{1}{4}}\delta_{\dalpha}{}^{\dbeta}\delta_{\dgamma}{}^{\ddelta}=\delta_{\dalpha}{}^{\ddelta}\delta_{\dgamma}{}^{\dbeta}\,. \label{6dfierz}
\end{equation}
Similarly for $\mbox{su}(2)$ indices we have
\begin{equation}
\textstyle{-\frac{1}{4}}(\sigma^{ij})_{\alpha}{}^{\beta}(\sigma_{ij})_{\gamma}{}^{\delta}+
\textstyle{\frac{1}{2}}\delta_{\alpha}{}^{\beta}\delta_{\gamma}{}^{\delta}=\delta_{\alpha}{}^{\delta}\delta_{\gamma}{}^{\beta}\,. \label{3dfierz}
\end{equation}

The nine dimensional charge conjugation matrix, ${C}$ is then given by, with $(\sigma^{i})^{T}=-\epsilon^{-1}\sigma^{i}\epsilon$, $A=1,2,\cdots,9$,
\begin{equation}
\begin{array}{cc}
(\Gamma^{A}){}^{T}=(\Gamma^{A}){}^{\ast}={C}^{-1}\Gamma^{A}{C}\,,~~~&~~~{C}=\epsilon\otimes\left(\begin{array}{cc}0&-1\\1&0\end{array}\right)\,,
\end{array}
\end{equation}
so that Majorana spinor satisfying  $\Psi={C}\Psi^{\ast}$  contains  eight independent  complex components
\begin{equation}
\begin{array}{cc}
\Psi=\left(\begin{array}{c} \psi_{\alpha\dalpha}\\ \\
\tilde{\psi}_{\alpha}{}^{\dalpha}\end{array}\right)\,,~~~&~~\tilde{\psi}_{\alpha}{}^{\dalpha}=\epsilon_{\alpha\beta}(\psi^{\ast})^{\beta\dalpha}\,.
\label{Majorana}
\end{array}
\end{equation}

\subsection{The Special Unitary Lie Superalgebra, $\mbox{su}(2|4\,;2,0)$ or $\mbox{su}(2|4\,;2,4)$}
In terms of the three and six dimensional gamma matrices, the superalgebra of the M-theory in a fully supersymmetric pp-wave background
reads\footnote{Here we disregard the possible inclusion of the central charges which may appear in the large $N$ limit of the matrix theory
\cite{Hyun:2002cm}.}
\begin{equation}
\begin{array}{cc}
\multicolumn{2}{c}{[H,Q_{\alpha\dalpha}]=\textstyle{\frac{\mu}{12}} Q_{\alpha\dalpha}\,,}\\
{}&{}\\
{}[M_{ij},Q_{\alpha\dalpha}]=i\textstyle{\frac{1}{2}}(\sigma_{ij})_{\alpha}{}^{\beta}Q_{\beta\dalpha}\,,~~~~~&~~~
{}[M_{ab},Q_{\alpha\dalpha}]=i\textstyle{\frac{1}{2}}(\rho_{ab})_{\dalpha}{}^{\dbeta}Q_{\alpha\dbeta}\,,~~\\
{}&{}\\
{}[M_{ij},\bar{Q}^{\alpha\dalpha}]=-i\textstyle{\frac{1}{2}}\bar{Q}^{\beta\dalpha}(\sigma_{ij})_{\beta}{}^{\alpha}\,,~~~&~~~
{}[M_{ab},\bar{Q}^{\alpha\dalpha}]=-i\textstyle{\frac{1}{2}}\bar{Q}^{\alpha\dbeta}(\rho_{ab})_{\dbeta}{}^{\dalpha}\,,\\
{}&{}\\
{}[M_{i},M_{j}]=i\epsilon_{ijk}M_{k}\,, & M_{i}=\textstyle{\frac{1}{2}}\epsilon_{ijk}M_{jk}\,,\\
{}&{}\\
\multicolumn{2}{c}{ [M_{ab},M_{cd}]=i(\delta_{ac}M_{bd}-\delta_{ad}M_{bc}-\delta_{bc}M_{ad}+\delta_{bd}M_{ac})\,,}\\
{}&{}\\
\multicolumn{2}{c}{\{Q_{\alpha\dalpha},\,\bar{Q}^{\beta\dbeta}\}=\delta_{\alpha}{}^{\beta}\delta_{\dalpha}{}^{\dbeta}H+i\textstyle{\frac{\mu}{6}}
(\sigma^{ij})_{\alpha}{}^{\beta}\delta_{\dalpha}{}^{\dbeta}M_{ij}-i\textstyle{\frac{\mu}{12}}
\delta_{\alpha}{}^{\beta}(\rho^{ab})_{\dalpha}{}^{\dbeta}M_{ab}\,.}\\
\end{array}
\label{susyalge}
\end{equation}
The sign difference   for the $\mbox{so}(3)$ and $\mbox{so}(6)$ generators appearing on the right hand side of the last line   is crucial for the
consistency. Along with $H$ the $\mbox{so}(3)$ and $\mbox{so}(6)$ generators  are hermitian, $(M_{ij})^{\dagger}=M_{ij}$, $(M_{ab})^{\dagger}=M_{ab}$.
Note also that  $\mbox{su}(2)\equiv\mbox{so}(3)$, $\mbox{su}(4)\equiv\mbox{so}(6)$.

In order to identify the superalgebra in terms of the supermatrices \cite{jhp}, we consider
\begin{equation}
{\cal K}{\cdot{\cal P}}= \phi H+\bar{\theta}^{\alpha\dalpha}Q_{\alpha\dalpha}+\bar{Q}^{\alpha\dalpha}\theta_{\alpha\dalpha}+
\textstyle{\frac{1}{2}}w^{ij}M_{ij}+\textstyle{\frac{1}{2}}w^{ab}M_{ab}
\end{equation}
where $\theta_{\alpha\dalpha},,\bar{\theta}^{\alpha\dalpha}=(\theta_{\alpha\dalpha})^{\dagger}$ are Grassmannian ``odd'' coordinates and
\begin{equation}
\begin{array}{cc}
{\cal K}=(\phi,\theta,\bar{\theta},w^{ij},w^{ab})\,,~~~&~~ {\cal P}=(H,\bar{Q},Q,M_{ij},M_{ab})\,.
\end{array}
\end{equation}

Setting
\begin{equation}
[{\cal K}_{1}{\cdot{\cal P}},{\cal K}_{2}{\cdot{\cal P}}] =-i{\cal K}_{3}{\cdot{\cal P}}\,,
\end{equation}
we get
\begin{equation}
\begin{array}{l}
\phi_{3}=i(\bar{\theta}_{1}\theta_{2}-\bar{\theta}_{2}\theta_{1})\,,\\
{}\\
\theta_{3}=\textstyle{\frac{1}{4}}w_{1}^{ij}\sigma_{ij}\,\theta_{2}+
\textstyle{\frac{1}{4}}w_{1}^{ab}\rho_{ab}\,\theta_{2}+i\textstyle{\frac{\mu}{12}}\phi_{2}\theta_{1}\,-\,(1\leftrightarrow 2)\,,\\
{}\\
w_{3}^{ij}=w_{1}^{i}{}_{k}w_{2}^{kj}+\textstyle{\frac{\mu}{3}}\bar{\theta}_{2}\sigma^{ij}\theta_{1}\,-\,(1\leftrightarrow 2)\,,\\
{}\\
w_{3}^{ab}=w_{1}^{a}{}_{c}w_{2}^{cb}-\textstyle{\frac{\mu}{6}}\bar{\theta}_{2}\rho^{ab}\theta_{1}\,-\,(1\leftrightarrow 2)\,. \label{3s}
\end{array}
\end{equation}

Now if we define a $(2+4)\times(2+4)$ supermatrix, ${\cal M}$, as
\begin{equation}
{\cal M}=\left(\begin{array}{cc} \left(\textstyle{\frac{1}{4}}w^{ij}\sigma_{ij}-i\textstyle{\frac{\mu}{6}}\phi\right)_{\alpha}{}^{\beta}~&~
\sqrt{\frac{\mu}{3}}\,\theta_{\alpha\dalpha}\\
{}&{}\\
\sqrt{\frac{\mu}{3}}\,\bar{\theta}^{\beta\dbeta}~&~
\left(\textstyle{\frac{1}{4}}w^{ab}(-\rho_{ab})^{T}-i\textstyle{\frac{\mu}{12}}\phi\right)^{\dbeta}{}_{\dalpha}
\end{array}
\right)\,,
\end{equation}
we can show using (\ref{6dfierz}) and (\ref{3dfierz})  that the relation above (\ref{3s}) agrees with the matrix commutator
\begin{equation}
[{\cal M}_{1}, {\cal M}_{2}]={\cal M}_{3}\,.
\end{equation}
In general, ${\cal M}$ can be defined as a $(2+4)\times(2+4)$ supermatrix subject to
\begin{equation}
\begin{array}{cc}
\mbox{str}\,{\cal M}=0\,,~~~&~~~-{\cal M}^{\dagger}=\left(\begin{array}{cc} 1&0\\
0&-\frac{\mu}{|\mu|}\end{array}\right){\cal M}\left(\begin{array}{cc} 1&0\\
0&-\frac{\mu}{|\mu|}\end{array}\right)\,.
\end{array}
\end{equation}
Thus, the superalgebra of the M-theory on a fully supersymmetric pp-wave is  the special unitary Lie superalgebra, $\mbox{su}(2|4\,;2,0)$ for $\mu>0$ or
$\mbox{su}(2|4\,;2,4)$ for $\mu<0$ having dimensions $(19|16)$. Our convention here is from  Kac~\cite{Kac:qb} such that the bosonic part of
$\mbox{su}(p|q;p^{\prime},q^{\prime})$ is given by $\mbox{su}(p^{\prime},p-p^{\prime})\oplus\mbox{su}(q^{\prime},q-q^{\prime})$.

 The fact that there are two inequivalent superalgebras for different signs of $\mu$  is essentially  due to the fact that in nine or odd dimensions
 there are two inequivalent classes of gamma matrices characterized by $\Gamma^{12\cdots 9}=1$ or $-1$
which are not related by the similarity transformations. Rewriting the superalgebra (\ref{susyalge}) in terms of the charge conjugates as in
(\ref{Majorana}), $S_{\alpha}{}^{\dalpha}\equiv\epsilon_{\alpha\beta}\bar{Q}^{\beta\dalpha}$, one can obtain the same superalgebra for the opposite sign
of $\mu$ with $\rho_{ab}$ replaced by $\bar{\rho}_{ab}\equiv \frac{1}{2}(\bar{\rho}_{a}\rho_{b}-\bar{\rho}_{b}\rho_{a})$,
\begin{equation}
\begin{array}{ccc}
\left(\begin{array}{c}
Q_{\alpha\dalpha}\\
{}\\
\mu\\
{}\\
(\rho_{ab})_{\dalpha}{}^{\dbeta}
\end{array}
\right)~~~~&~~~\longleftrightarrow~~~&~~~~
\left(\begin{array}{c}
S_{\alpha}{}^{\dalpha}~\\
{}\\
-\mu~~\\
{}\\
(\bar{\rho}_{ab})^{\dalpha}{}_{\dbeta}
\end{array}
\right)\,.
\end{array}
\end{equation}
The exchange of $\rho_{a}\leftrightarrow\bar{\rho}_{a}$ corresponds to the inequivalent choices of gamma matrices, $\Gamma^{12\cdots 9}=1$ or $-1$.
Therefore, sticking to one class of gamma matrices, but allowing both $+$ and $-$ signs for $\mu$ we do not lose any generality. Our choice is from
(\ref{gamma7})  $\Gamma^{12\cdots 9}=1$. Nevertheless after all, we will see that the physics is independent of the sign.

\section{The Root Structure of the Superalgebra\label{root}}
In this section we analyze the root structure  of the superalgebra. We first start with the following $8\times 8$ representation of the bosonic part,
$\mbox{u}(1)\oplus\mbox{su}(2)\oplus\mbox{su}(4)$
\begin{equation}
\Big(R(\textstyle{\frac{6}{\mu}}H)\,,~R(M_{ij})\,,~R(M_{ab})\Big)=\Big(-\textstyle{\frac{1}{2}}\otimes 1\,,~-i\textstyle{\frac{1}{2}}\sigma_{ij}\otimes
1\,,~1\otimes -i\textstyle{\frac{1}{2}}\rho_{ab}\Big)\,,
\end{equation}
which are orthonormal and hermitian
\begin{equation}
\begin{array}{ccc}
\Tr(R_{I}^{\dagger}R_{J})=2\delta_{IJ}\,,~~~&~~~R_{I}^{\dagger}=R_{I}\,,~~~&~~~I,J=1,2,\cdots,19\,.
\end{array}
\label{RI}
\end{equation}
Our choice of the Cartan subalgebra is
\begin{equation}
\H=(\textstyle{\frac{6}{\mu}}H,\,M_{12},\,M_{45},\,M_{67},\, M_{89})\,.
\end{equation}
Using the $\mbox{U}(4)$ symmetry, $\rho_{a}\rightarrow U\rho_{a}U^{T}$, $UU^{\dagger}=1$, which preserves the anti-symmetric property (\ref{anti-sym}) of
$\rho_{a}$, we can take the representation of the Cartan subalgebra in a diagonal form. Adopting the bra and ket notation for the $\mbox{su}(4)$ part we
set
\begin{equation}
\begin{array}{l}
R(\textstyle{\frac{6}{\mu}}H)=-\textstyle{\frac{1}{2}}\otimes 1\,,~~~~~~~~~
R(M_{12})=\textstyle{\frac{1}{2}}\sigma_{3}\otimes 1\,,\\
{}\\
R(M_{45})=1\otimes \textstyle{\frac{1}{2}}\left(\,|1\rangle\langle 1|+|2\rangle\langle 2|-|3\rangle\langle 3|-|4\rangle\langle
4|\,\right)\,,\\
{}\\
R(M_{67})=1\otimes \textstyle{\frac{1}{2}}\left(\,|1\rangle\langle 1|-|2\rangle\langle 2|+|3\rangle\langle 3|-|4\rangle\langle
4|\,\right)\,,\\
{}\\
R(M_{89})=1\otimes \textstyle{\frac{1}{2}}\left(\,|1\rangle\langle 1|-|2\rangle\langle 2|-|3\rangle\langle 3|+|4\rangle\langle 4|\,\right)\,,
\end{array}
\label{cartan}
\end{equation}
which is also compatible with (\ref{gamma7}).\\

All the  bosonic positive roots are then given by
\begin{equation}
\begin{array}{ll}
R(\E_{z})=\textstyle{\frac{1}{\sqrt{2}}}\left(\begin{array}{cc}0&1\\0&0\end{array}\right)\otimes 1\,,~~&z=(0,1,0,0,0)\,,\\
{}&{}\\
R(\E_{u})=1\otimes |1\rangle\langle 2|\,,~~&u=(0,0,0,1,1)\,,\\
{}&{}\\
R(\E_{v})=1\otimes |2\rangle\langle 3|\,,~~&v=(0,0,1,-1,0)\,,\\
{}&{}\\
R(\E_{w})=1\otimes |3\rangle\langle 4|\,,~~&w=(0,0,0,1,-1)\,,\\
{}&{}\\
R(\E_{u+v})=1\otimes |1\rangle\langle 3|\,,~~&u+v=(0,0,1,0,1)\,,\\
{}&{}\\
R(\E_{v+w})=1\otimes |2\rangle\langle 4|\,,~~&v+w=(0,0,1,0,-1)\,,\\
{}&{}\\
R(\E_{u+v+w})=1\otimes |1\rangle\langle 4|\,,&u+v+w=(0,0,1,1,0)\,,
\end{array}
\end{equation}
where $z$ and $u,v,w$ are respectively the $\mbox{su}(2)$ and $\mbox{su}(4)$ simple roots. The negative roots follow  simply from
$R(\E_{-z})=R(\E_{z})^{\dagger}\,,$ etc.

Just like $R_{I}$ in (\ref{RI}), $(R(\H),\,R(\E_{+}),\,R(\E_{-}))$ are also orthonormal. This implies that those two are related by the unitary
transformation so that
\begin{equation}
\begin{array}{ll}
\textstyle{\frac{1}{2}}R(M^{ab})M_{ab} =&R(M_{45})M_{45}+R(M_{67})M_{67}+R(M_{89})M_{89}\\{}&{}\\
{}&\, + \displaystyle{\sum_{\chi\in\Delta_{4}^{+}}}\, \Big(R(\E_{\chi})\E_{-\chi}+R(\E_{-\chi})\E_{\chi}\Big)\,,
\end{array}
\label{MR}
\end{equation}
where $\Delta_{4}^{+}$ denotes the set of all the $\mbox{su}(4)$ positive roots $u,\,v,\,w,\,u+v,\,v+w,\,u+v+w$.

In terms of the Cartan subalgebra and any $\mbox{su}(2)\oplus\mbox{su}(4)$ root, $\chi$,  the  superalgebra of the M-theory on a pp-wave reads  up to the
complex conjugate
\begin{equation}
\begin{array}{ll}
{}[\H,\E_{\chi}]=\chi \E_{\chi}\,,~~~&~~~ [\E_{\chi},\E_{-\chi}]=\chi{\cdot\H}\,,\\
{}&{}\\
{}[\E_{u},\E_{v}]=\E_{u+v}\,,~~~&~~~[\E_{v},\E_{w}]=\E_{v+w}\,,\\
{}&{}\\
{}[\E_{u+v},\E_{w}]=\E_{u+v+w}\,,~~~&~~~[\E_{u},\E_{w}]=0\,,\\
{}&{}\\
{}[\H,Q]=-R(\H)Q\,,~~~&~~~[\E_{\chi},Q]=-R(\E_{\chi})Q\,,\label{Qalge}
\end{array}
\end{equation}
and
\begin{equation}
\begin{array}{ll}
\{Q_{\alpha\dalpha},\bar{Q}^{\beta\dbeta}\}=&
\delta_{\alpha}{}^{\beta}\delta_{\dalpha}{}^{\dbeta}H-\textstyle{\frac{\mu}{3}}\left(\begin{array}{cc} M_{12}&\sqrt{2}\E_{-z}\\
\sqrt{2}\E_{z}&-M_{12}\end{array}\right)_{\alpha}^{~~\beta}\,{}\delta_{\dalpha}{}^{\dbeta}\\
{}&{}\\
{}&\,+\textstyle{\frac{\mu}{3}}\delta_{\alpha}{}^{\beta}\left(\begin{array}{cccc}
f_{1}&\E_{-u}&\E_{-u-v}&\E_{-u-v-w}\\
\E_{u}&f_{2}&\E_{-v}&\E_{-v-w}\\
\E_{u+v}&\E_{v}&f_{3}&\E_{-w}\\
\E_{u+v+w}&\E_{v+w}&\E_{w}&f_{4}\end{array}\right)_{\dalpha}^{~~\dbeta}\,,
\end{array}
\label{QQ}
\end{equation}
where
\begin{equation}
\begin{array}{ll}
f_{1}=\frac{1}{2}(M_{45}+M_{67}+M_{89})\,,~~~&~~~f_{2}=\frac{1}{2}(M_{45}-M_{67}-M_{89})\,,\\
{}&{}\\
f_{3}=\frac{1}{2}(-M_{45}+M_{67}-M_{89})\,,~~~&~~~f_{4}=\frac{1}{2}(-M_{45}-M_{67}+M_{89})\,.
\end{array}
\end{equation}

In particular from (\ref{Qalge}), the unique fermionic simple root, $q$ and other fermionic positive roots are given as
\begin{equation}
\begin{array}{ll}
{}[\H,Q_{11}]=qQ_{11}\,,~~~&~~q=(\textstyle{\frac{1}{2},-\frac{1}{2},-\frac{1}{2},-\frac{1}{2},-\frac{1}{2}})\,,\\
{}&{}\\
{}[\H,Q_{12}]=(q+u)Q_{12}\,,~~~&~~Q_{12}=-[\E_{u},Q_{11}]\,,\\
{}&{}\\
{}[\H,Q_{13}]=(q+u+v)Q_{13}\,,~~~&~~Q_{13}=-[\E_{u+v},Q_{11}]\,,\\
{}&{}\\
{}[\H,Q_{14}]=(q+u+v+w)Q_{14}\,,~~~&~~Q_{14}=-[\E_{u+v+w},Q_{11}]\,,\\
{}&{}\\
{}[\H,Q_{21}]=(q+z)Q_{21}\,,~~~&~~Q_{21}=-[\sqrt{2}\E_{z},Q_{11}]\,,\\
{}&{}\\
{}[\H,Q_{22}]=(q+z+u)Q_{22}\,,~~~&~~Q_{12}=-[\sqrt{2}\E_{z},Q_{12}]\,,\\
{}&{}\\
{}[\H,Q_{23}]=(q+z+u+v)Q_{23}\,,~~~&~~Q_{23}=-[\sqrt{2}\E_{z},Q_{13}]\,,\\
{}&{}\\
{}[\H,Q_{24}]=(q+z+u+v+w)Q_{24}\,,~~~&~~Q_{24}=-[\sqrt{2}\E_{z},Q_{14}]\,.
\end{array}
\label{qQ}
\end{equation}
\newpage

The complexification of $\mbox{su}(2|4\,;2,0)$ or $\mbox{su}(2|4\,;2,4)$ corresponds to $\mbox{A}(1|3)$, and its Dynkin diagram is with the simple roots

\begin{figure}[!htb]
\vspace{15mm}
\begin{center}
~\epsfig{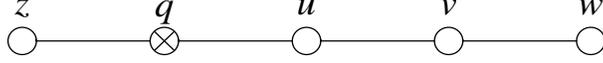}
\end{center}
\caption{The Dynkin diagram of $\mbox{A}(1|3)$} \label{f1} \vspace{15mm}
\end{figure}

Finally the second order Casimir operator, ${\cal C}_{{\scriptstyle M}}$ is
\begin{equation}
{\cal C}_{{\scriptstyle M}}=\textstyle{\frac{12}{\mu}}H^{2}-\textstyle{\frac{\mu}{3}}{\cal C}_{4} +\textstyle{\frac{2\mu}{3}}{\cal
C}_{2}+\bar{Q}^{\alpha\dalpha}Q_{\alpha\dalpha}-Q_{\alpha\dalpha}\bar{Q}^{\alpha\dalpha}\,,
\end{equation}
where ${\cal C}_{4}$ and ${\cal C}_{2}$ are $\mbox{su}(4)$ and $\mbox{su}(2)$ Casimirs
\begin{equation}
\begin{array}{l}
{\cal C}_{4}=\textstyle{\frac{1}{2}}M^{ab}M_{ab}=M_{45}^{2}+M_{67}^{2}+M_{89}^{2}+\displaystyle{\sum_{\chi\in\Delta^{+}_{4}}}\,
\Big(\E_{\chi}\E_{-\chi}+\E_{-\chi}\E_{\chi}\Big)\,,\\
{}\\
{\cal C}_{2}=M_{12}^{2}+\E_{z}\E_{-z}+\E_{-z}\E_{z}\,.
\end{array}
\end{equation}

\section{Irreducible Representations\label{IRR}}
Starting with an eigenstate of $H$, by acting $\bar{Q}^{\alpha\dalpha}$'s as many as possible maximally eight times surely,  one can obtain a state which
is annihilated by all the $\bar{Q}^{\alpha\dalpha}$'s.  Now under the action of the bosonic operators, the state opens up  an irreducible representation
of $\mbox{u}(1)\oplus\mbox{su}(2)\oplus\mbox{su}(4)$ or the zeroth floor multiplet. Further from (\ref{Qalge}), any state in the  multiplet  is
annihilated by all the fermionic negative roots. Consequently there exists a unique \textit{superlowest weight}, $|\Lambda_{L}\rangle$, in the
supermultiplet  annihilated by all the negative roots
\begin{equation}
\begin{array}{ll}
\bar{Q}^{11}|\Lambda_{L}\rangle=0\,,~~~&~~~\E_{-\chi}|\Lambda_{L}\rangle=0\,,~~\chi=z,u,v,w.
\end{array}
\end{equation}
The superlowest weight vector  is specified by an arbitrary energy value, $E_{0}$ and four non-negative integers, $J_{z},J_{u},J_{v},J_{w}$
\begin{equation}
\Lambda_{L}=\left(\textstyle{\frac{6}{\mu}E_{0},\,{-\frac{1}{2}J_{z}}\,,\,{-\frac{1}{2}(J_{u}+2J_{v}+J_{w})}\,,\,{
-\frac{1}{2}(J_{u}+J_{w})}\,,\,{-\frac{1}{2}(J_{u}-J_{w})}}\right) \label{Lambda}
\end{equation}
satisfying for the $\mbox{su}(2)\oplus\mbox{su}(4)$  simple roots, $\chi=z,u,v,w$,
\begin{equation}
\begin{array}{cc}
\displaystyle{-2\frac{\,\chi{\cdot\Lambda_{L}}\,}{\chi^{2}}=J_{\chi}}\,,~~~&~~~(\E_{\chi})^{J_{\chi}+1}|\Lambda_{L}\rangle=0\,.
\end{array}
\end{equation}

All the other states are generated by repeated applications of the positive roots  on $|\Lambda_{L}\rangle$, and without loss of generality one can
safely restrict on the simple roots only, $Q_{11}$, $\E_{\chi},\,\chi=z,u,v,w$. In general the different  ordering of multiplication may result in the
degeneracy for the states  of a definite weight vector. Using the commutator relations $[\E_{\chi},Q]\sim Q$ in (\ref{Qalge}), one can always move all
the $Q_{11}$'s appearing to either far right or far left allowing other fermionic positive roots.  Therefore the whole supermultiplet is exhibited  by
\begin{equation}
\E_{\chi_{n}}\cdots \E_{\chi_{1}}Q_{2\dbeta_{l}}\cdots Q_{2\dbeta_{1}}Q _{1\dalpha_{m}}\cdots Q_{1\dalpha_{1}}|\Lambda_{L}\rangle \label{LLQQ}
\end{equation}
or  equivalently
\begin{equation}
Q_{2\dbeta_{l}}\cdots Q_{2\dbeta_{1}}Q _{1\dalpha_{m}}\cdots Q_{1\dalpha_{1}}\E_{\chi_{n}}\cdots \E_{\chi_{1}}|\Lambda_{L}\rangle\,. \label{finite}
\end{equation}
The latter is essentially the repeated applications of the fermionic positive roots on  the zeroth floor.  As the zeroth floor multiplet has dimension
\cite{Humphreys:dw}
\begin{equation}
d_{0}=(J_{z}+1)\times\Big[\textstyle{\frac{1}{12}}(J_{u}+1)(J_{v}+1)(J_{w}+1)(J_{u}+J_{v}+2)(J_{v}+J_{w}+2)(J_{u}+J_{v}+J_{w}+3)\Big]\,, \label{bdim}
\end{equation}
Eq.(\ref{finite}) implies that the supermultiplet has a finite dimension, $d_{s}$
\begin{equation}
d_{s}\leq 256\times d_{0}\,.
\end{equation}

The application of a $Q_{\alpha\dalpha}$ changes the $\mbox{u}(1)\oplus\mbox{su}(2)\oplus\mbox{su}(4)$ multiplets jumping from an irreducible
representation to another. In particular, the number of the applied fermionic roots determines the floor number, zero to eight at most. Each floor has the
energy
\begin{equation}
\begin{array}{cc}
E_{N}=E_{0}+\frac{\mu}{12}N\,,~~~&~~~N=0,1,2,\cdots,8.
\end{array}
\end{equation}
Apparently the zeroth floor corresponds to the lowest energy  states  for $\mu>0$ or to the highest energy states for $\mu< 0$.

Each of the zeroth and the highest floors   corresponds to an  irreducible representation of $\mbox{u}(1)\oplus\mbox{su}(2)\oplus\mbox{su}(4)$, while
others are in general  reducible representations.   Each irreducible representation is  specified by its lowest weight, $\lambda_{L}$, annihilated by all
the bosonic negative roots of the form
\begin{equation}
\lambda_{L}=\left(\textstyle{\frac{6}{\mu}E_{0}+\frac{1}{2}N,\,{-\frac{1}{2}j_{z}}\,,\,{-\frac{1}{2}(j_{u}+2j_{v}+j_{w})}\,,\,{
-\frac{1}{2}(j_{u}+j_{w})}\,,\,{-\frac{1}{2}(j_{u}-j_{w})}}\right)\,.
\end{equation}
The corresponding highest weight is then \cite{Fulton}
\begin{equation}
\lambda_{H}=\left(\textstyle{\frac{6}{\mu}E_{0}+\frac{1}{2}N,\,{\frac{1}{2}j_{z}}\,,\,{\frac{1}{2}(j_{w}+2j_{v}+j_{u})}\,,\,{
\frac{1}{2}(j_{w}+j_{u})}\,,\,{\frac{1}{2}(j_{w}-j_{u})}}\right)\,,
\end{equation}
while the dimension is given by (\ref{bdim}) with $J\leftrightarrow j$.

 Now if we  define
\begin{equation}
\begin{array}{cc}
T^{l}_{\alpha}\equiv Q_{\alpha l}\cdots Q_{\alpha 2}Q_{\alpha 1}&~~~~~\mbox{for}~~l=1,2,3,4
\end{array}
\end{equation}
then  all the ``naturally ordered" products of the fermionic  roots are given by \cite{Hurni:1981ki}
\begin{equation}
\begin{array}{ccc}
T^{m}_{1}\,,~~~~&~~~~T^{l}_{2}{}T^{m}_{1}\,,~~~&~~1\leq l\leq m\leq 4\,.
\end{array}
\end{equation}
There are fourteen of them and they are  all the possible products of $Q_{\alpha\dalpha}$'s  which commute with any bosonic negative root
\begin{equation}
\begin{array}{ccc}
{}[\E_{-\chi},T^{m}_{1}]=0\,,~~~&~~~{}[\E_{-\chi},T^{l}_{2}{}T^{m}_{1}]=0\,,~~~&~~~1\leq l\leq m\leq 4\,.
\end{array}
\label{LTT}
\end{equation}
Apparently we have the following possible lowest weights for $\mbox{u}(1)\oplus\mbox{su}(2)\oplus\mbox{su}(4)$ multiplets on each floor
\begin{equation}
\begin{array}{lc}
N=8~:&T^{4}_{2}T^{4}_{1}|\Lambda_{L}\rangle\\
{}&{}\\
N=7~:&T^{3}_{2}T^{4}_{1}|\Lambda_{L}\rangle\\
{}&{}\\
N=6~:&T^{2}_{2}T^{4}_{1}|\Lambda_{L}\rangle\,,~~~~~~~~~~~~T^{3}_{2}T^{3}_{1}|\Lambda_{L}\rangle\\
{}&{}\\
N=5~:&T^{1}_{2}T^{4}_{1}|\Lambda_{L}\rangle\,,~~~~~~~~~~~~T^{2}_{2}T^{3}_{1}|\Lambda_{L}\rangle\\
{}&{}\\
N=4~:~~~~~~~&T^{4}_{1}|\Lambda_{L}\rangle\,,~~~~~~~~~~~~~~~~T^{1}_{2}T^{3}_{1}|\Lambda_{L}\rangle\,,~~~~~~~~~~~~T^{2}_{2}T^{2}_{1}|\Lambda_{L}\rangle\\
{}&{}\\
N=3~:&~~T^{3}_{1}|\Lambda_{L}\rangle\,,~~~~~~~~~~~~T^{1}_{2}T^{2}_{1}|\Lambda_{L}\rangle\\
{}&{}\\
N=2~:&~~T^{2}_{1}|\Lambda_{L}\rangle\,,~~~~~~~~~~~~T^{1}_{2}T^{1}_{1}|\Lambda_{L}\rangle\\
{}&{}\\
N=1~:& T^{1}_{1}|\Lambda_{L}\rangle\\
{}&{}\\
N=0~:&|\Lambda_{L}\rangle\,.
\end{array}
\label{lowestweights}
\end{equation}
A few remarks are in order. They are truly lowest weights only if they do not decouple. The dimension of the corresponding
$\mbox{u}(1)\oplus\mbox{su}(2)\oplus\mbox{su}(4)$ irreducible representation can be easily obtained from (\ref{qQ}), (\ref{Lambda}) and (\ref{bdim}). One
finds that the sum of the dimensions is equal to $256\times d_{0}$ only if the zeroth floor is a $\mbox{su}(2)\oplus\mbox{su}(4)$  singlet, i.e.
$d_{0}=1$, and  for the non-singlet zeroth floor it is strictly less. Thus in general, apart from (\ref{lowestweights}), there are hidden lowest weights,
of which the annihilation  by the bosonic negative roots requires   the explicit information of the superlowest weight vector (\ref{Lambda}). For
example, a possible hidden lowest weight is
\begin{equation}
(J_{u}Q_{12}+Q_{11}\E_{u})|\Lambda_{L}\rangle\,,
\end{equation}
which vanishes identically if $J_{u}=0$.

\subsection{Typical and Atypical Representations}
Typical representation is defined to have  the possible maximal dimension, $d_{s}=256\times d_{0}$, while  atypical representation has less dimension.
Namely, in a typical representation all the possible states which could appear will appear. Consequently for typical representation,
$T^{4}_{2}T^{4}_{1}|\Lambda_{L}\rangle$ does not decouple
\begin{equation}
T^{4}_{2}T^{4}_{1}|\Lambda_{L}\rangle\neq 0. \label{8Q}
\end{equation}
\textit{Proposition}~: An irreducible representation is typical if and only if  Eq.(\ref{8Q}) holds.\\
\textit{Proof}~: We only need to prove the sufficiency. First  consider an arbitrary state on the zeroth floor, $|\lambda\rangle$, and the corresponding
bosonic operator, ${\cal B}_{-}$ which  is essentially a product of bosonic negative roots taking $|\lambda\rangle$ to the superlowest weight
\begin{equation}
{\cal B}_{-}|\lambda\rangle=|\Lambda_{L}\rangle\,.
\end{equation}
Now consider any state, $|\upsilon\rangle$ in the supermultiplet and write it as (\ref{finite})
\begin{equation}
|\upsilon\rangle=Q_{2\dbeta_{l}}\cdots Q_{2\dbeta_{1}}Q _{1\dalpha_{m}}\cdots Q_{1\dalpha_{1}}|\lambda\rangle\,.
\end{equation}
Acting the complementary fermionic positive roots and ${\cal B}_{-}$ successively  we obtain using (\ref{LTT})
\begin{equation}
{\cal B}_{-}Q_{2\dbeta_{4}}\cdots Q_{2\dbeta_{l+1}}Q _{1\dalpha_{4}}\cdots Q_{1\dalpha_{m+1}}|\upsilon\rangle ={\cal
B}_{-}T^{4}_{2}T^{4}_{1}|\lambda\rangle=T^{4}_{2}T^{4}_{1}|\Lambda_{L}\rangle\,.
\end{equation}
Thus (\ref{8Q}) implies the nonvanishing of $|\upsilon\rangle$.  This completes our proof.\newline

To see whether  $T^{4}_{2}T^{4}_{1}|\Lambda_{L}\rangle$ decouples or not  we need the following recurrent relations
\begin{equation}
\begin{array}{l}
(T_{1}^{l})^{\dagger}T_{1}^{l}|\Lambda_{L}\rangle=c_{l}(T_{1}^{l-1})^{\dagger}T_{1}^{l-1}|\Lambda_{L}\rangle\,,\\
{}\\
(T_{2}^{l}T_{1}^{m})^{\dagger}T_{2}^{l}T^{m}_{1}|\Lambda_{L}\rangle=c_{l+4}(T_{2}^{l-1}T_{1}^{m})^{\dagger}T_{2}^{l-1}T_{1}^{m}|\Lambda_{L}\rangle\,,
\end{array}
\end{equation}
where   $1\leq l\leq m\leq 4$, $~T_{\alpha}^{0}=1$  and $c_{l+4}$ is independent of $m$
\begin{equation}
\begin{array}{l}
c_{1}=E_{0}+\textstyle{\frac{\mu}{12}}(2J_{z}-3J_{u}-2J_{v}-J_{w})\,,\\
{}\\
c_{2}=E_{0}+\textstyle{\frac{\mu}{12}}(2J_{z}+J_{u}-2J_{v}-J_{w}+4)\,,\\
{}\\
c_{3}=E_{0}+\textstyle{\frac{\mu}{12}}(2J_{z}+J_{u}+2J_{v}-J_{w}+8)\,,\\
{}\\
c_{4}=E_{0}+\textstyle{\frac{\mu}{12}}(2J_{z}+J_{u}+2J_{v}+3J_{w}+12)\,,\\
{}\\
c_{5}=E_{0}-\textstyle{\frac{\mu}{12}}(2J_{z}+3J_{u}+2J_{v}+J_{w}+4)\,,\\
{}\\
c_{6}=E_{0}-\textstyle{\frac{\mu}{12}}(2J_{z}-J_{u}+2J_{v}+J_{w})\,,\\
{}\\
c_{7}=E_{0}-\textstyle{\frac{\mu}{12}}(2J_{z}-J_{u}-2J_{v}+J_{w}-4)\,,\\
{}\\
c_{8}=E_{0}-\textstyle{\frac{\mu}{12}}(2J_{z}-J_{u}-2J_{v}-3J_{w}-8)\,.
\end{array}
\label{atypical}
\end{equation}
Consequently
\begin{equation}
(T_{2}^{4}T_{1}^{4})^{\dagger}T_{2}^{4}T^{4}_{1}|\Lambda_{L}\rangle=\prod_{k=1}^{8}c_{k}|\Lambda_{L}\rangle\,.
\end{equation}
Thus  for all $k=1,2,\cdots,8$  when $c_{k}\neq 0$,~  $T_{2}^{4}T^{4}_{1}|\Lambda_{L}\rangle$ can not decouple and the supermultiplet is typical.
Otherwise the representation is shortened or atypical and the energy value is quantized.

Similarly it is also interesting to note that if $T^{4}_{1}|\Lambda_{L}\rangle\neq 0$, then all the states of the form
\begin{equation}
\begin{array}{cc}
Q _{1\dalpha_{m}}\cdots Q_{1\dalpha_{1}}|\lambda\rangle\,,~~~&~~~0\leq m\leq 4
\end{array}
\end{equation}
do not decouple, where   $|\lambda\rangle$ is an arbitrary state of the zeroth floor.

\subsection{BPS Multiplets}
BPS multiplets are naturally associated to a superspace with lower number of ``odd'' coordinates \cite{Ferrara:2000dv}. Namely they are  a kind of
atypical unitary representation of which \textit{either} the superlowest weight is annihilated by a certain number of  supercharges,
$Q_{\alpha\dalpha}|\Lambda_{L}\rangle=0$ \textit{or} the superhighest weight is so, $\bar{Q}^{\alpha\dalpha}|\Lambda_{H}\rangle=0$. First we focus on the
case for the superlowest weight. Acting the bosonic negative roots we further get
\begin{equation}
\begin{array}{cc}
Q_{\beta\dbeta}|\Lambda_{L}\rangle=0~~~&~\mbox{for}~~1\leq\beta\leq\alpha\,,~~~1\leq\dbeta\leq\dalpha\,.
\end{array}
\label{QBPS}
\end{equation}
Hence the constraint also removes  $Q_{\beta\dbeta}$, $\beta\leq\alpha,$ $\dbeta\leq\dalpha$ completely from the supermultiplet written in the form
(\ref{LLQQ}) implying the elimination  of the  corresponding ``odd'' coordinates.  By unitarity we mean the positive definite norm,  the physically
relevant condition. For generic $Q_{\beta\dbeta}$
\begin{equation}
\begin{array}{ll}
||Q_{\beta\dbeta}|\Lambda_{L}\rangle||^{2}=&E_{0}+\textstyle{\frac{\mu}{6}}J_{z}(\sigma_{3})_{\beta}{}^{\beta}\\
{}&{}\\
{}&\!\!+\textstyle{\frac{\mu}{12}}\mbox{diag}(-3J_{u}-2J_{v}-J_{w},\,J_{u}-2J_{v}-J_{w},\,J_{u}+2J_{v}-J_{w},\,J_{u}+2J_{v}+3J_{w})_{\dbeta}{}^{\dbeta}\,,
\label{QLL}
\end{array}
\end{equation}
so that
\begin{equation}
\begin{array}{ll}
E_{0}\geq \textstyle{\frac{\mu}{12}}(2J_{z}+3J_{u}+2J_{v}+J_{w})~~~&~~~\mbox{for~~}\mu>0\,, \\{}&{}\\
E_{0}\geq \textstyle{\frac{|\mu|}{12}}(2J_{z}+J_{u}+2J_{v}+3J_{w})~~~&~~~\mbox{for~~}\mu<0\,.
\end{array}
\end{equation}
The saturation, if ever happens,  must occur at least with $Q_{21}$ for $\mu>0$ and  $Q_{14}$ for $\mu<0$.

Combining the results above we get the following classification of the BPS multiplets.\\

{\bf $\mbox{SU}(2)$  Singlet BPS}  $~~(\mu>0)$\\

\begin{tabular}{lllll}
$4/16$ BPS : &$E_{0}=\frac{\mu}{12}(3J_{u}+2J_{v}+J_{w})$,& $J_{z}=0$, $J_{u}\neq 0$, &$(\alpha,\dalpha)=(2,1)$\\
{}&{}&{}&{}&{}\\
$8/16$ BPS : &$E_{0}=\frac{\mu}{12}(2J_{v}+J_{w})$, &$J_{z}=J_{u}= 0$, $J_{v}\neq 0$, &$(\alpha,\dalpha)=(2,2)$\\
{}&{}&{}&{}&{}\\
$12/16$ BPS : &$E_{0}=\frac{\mu}{12}J_{w}$,& $J_{z}=J_{u}=J_{v}=0$, $J_{w}\neq 0$, &$(\alpha,\dalpha)=(2,3)$\\
{}&{}&{}&{}&{}\\
Vacua  : &$E_{0}=0$,&$J_{z}=J_{u}=J_{v}=J_{w}=0$,& $(\alpha,\dalpha)=(2,4)$
\end{tabular}\\
\\
\\

{\bf $\mbox{SU}(4)$  Singlet BPS}  $~~(\mu<0)$\\

\begin{tabular}{lllll}
$8/16$ BPS : &$E_{0}=\frac{|\mu|}{6}J_{z},~~~~~~~~~~~~~~~~~$& $J_{u}=J_{v}=J_{w}= 0$, $J_{z}\neq 0$, &~$(\alpha,\dalpha)=(1,4)$\\
{}&{}&{}&{}&{}\\
Vacua : &$E_{0}=0$,&$J_{z}=J_{u}=J_{v}=J_{w}=0$,& $~(\alpha,\dalpha)=(2,4)$
\end{tabular}\\

\noindent where $(\alpha,\dalpha)$ refers the highest fermionic root in (\ref{QBPS}).

Although  it appears here that the sign of $\mu$ would matter, it is not completely asymmetric, since instead  of the superlowest weight one can
construct the BPS multiplet from the \textit{superhighest weight}, $|\Lambda_{H}\rangle$, to obtain the exactly same BPS spectrum as above but for the
opposite sign. More explicitly
\begin{eqnarray}
&\Lambda_{H}=\left(\textstyle{\frac{6}{\mu}E_{H},\,{\frac{1}{2}J^{\prime}_{z}}\,,\,{\frac{1}{2}(J^{\prime}_{u}+2J^{\prime}_{v}+J^{\prime}_{w})}\,,\,{
\frac{1}{2}(J^{\prime}_{u}+J^{\prime}_{w})}\,,\,{\frac{1}{2}(J^{\prime}_{u}-J^{\prime}_{w})}}\right)\,,\\
{}\nonumber\\
&\begin{array}{llll} \bar{Q}^{\alpha\dalpha}|\Lambda_{H}\rangle=0~~~&\Longrightarrow~~~&\bar{Q}^{\beta\dbeta}|\Lambda_{H}\rangle=0
~&~\mbox{for}~~1\leq\beta\leq\alpha\,,~~~1\leq\dbeta\leq\dalpha\,,
\end{array}
\end{eqnarray}
and
\begin{equation}
\begin{array}{ll}
||\bar{Q}^{\beta\dbeta}|\Lambda_{H}\rangle||^{2}=&E_{H}-\textstyle{\frac{\mu}{6}}J^{\prime}_{z}(\sigma_{3})_{\beta}{}^{\beta}\\
{}&{}\\
{}&\!\!-\textstyle{\frac{\mu}{12}}\mbox{diag}(-3J^{\prime}_{u}-2J^{\prime}_{v}-J^{\prime}_{w},\,
J^{\prime}_{u}-2J^{\prime}_{v}-J^{\prime}_{w},\,J^{\prime}_{u}+2J^{\prime}_{v}-J^{\prime}_{w},\,
J^{\prime}_{u}+2J^{\prime}_{v}+3J^{\prime}_{w})_{\dbeta}{}^{\dbeta}\,,
\end{array}
\end{equation}
where, compared to (\ref{QLL}),  the sign of $\mu$ is flipped while $J$ replaced by $J^{\prime}$.
The BPS multiplets and the energy spectrum are identical to the previous result\\

{\bf $\mbox{SU}(2)$  Singlet BPS}  $~~(\mu<0)$\\

\begin{tabular}{lllll}
$4/16$ BPS : &$E_{H}=\frac{|\mu|}{12}(3J^{\prime}_{u}+2J^{\prime}_{v}+J^{\prime}_{w})$,& $J^{\prime}_{z}=0$, $J^{\prime}_{u}\neq 0$, &$(\alpha,\dalpha)=(2,1)$\\
{}&{}&{}&{}&{}\\
$8/16$ BPS : &$E_{H}=\frac{|\mu|}{12}(2J^{\prime}_{v}+J^{\prime}_{w})$, &$J^{\prime}_{z}=J^{\prime}_{u}= 0$, $J^{\prime}_{v}\neq 0$, &$(\alpha,\dalpha)=(2,2)$\\
{}&{}&{}&{}&{}\\
$12/16$ BPS : &$E_{H}=\frac{|\mu|}{12}J^{\prime}_{w}$,& $J^{\prime}_{z}=J^{\prime}_{u}=J^{\prime}_{v}=0$, $J^{\prime}_{w}\neq 0$, &$(\alpha,\dalpha)=(2,3)$\\
{}&{}&{}&{}&{}\\
Vacua : &$E_{H}=0$,&$J^{\prime}_{z}=J^{\prime}_{u}=J^{\prime}_{v}=J^{\prime}_{w}=0$,& $(\alpha,\dalpha)=(2,4)$
\end{tabular}\\
\\
\\

{\bf $\mbox{SU}(4)$  Singlet BPS}  $~~(\mu>0)$\\

\begin{tabular}{lllll}
$8/16$ BPS : &$E_{H}=\textstyle{\frac{\mu}{6}}J^{\prime}_{z},~~~~~~~~~~~~~~~~~$&~$J^{\prime}_{u}=J^{\prime}_{v}=J^{\prime}_{w}= 0$,
$J^{\prime}_{z}\neq 0$, &~$(\alpha,\dalpha)=(1,4)$\\
{}&{}&{}&{}&{}\\
Vacua : &$E_{H}=0$,&~$J^{\prime}_{z}=J^{\prime}_{u}=J^{\prime}_{v}=J^{\prime}_{w}=0$,&~$(\alpha,\dalpha)=(2,4)$
\end{tabular}\\

In summary, there are two ways of defining the BPS multiplets, one from the superlowest weight and the other from the superhighest weight. Whichever we
choose, if it corresponds to the lowest  energy, the ``lowest energy''  floor is a $\mbox{su}(2)$ singlet, while if it corresponds to the highest energy,
the ``highest energy'' floor is a $\mbox{su}(4)$ singlet. After all the physics is independent of the sign of $\mu$.

It is worth noting that the $\mbox{su}(2)$ singlet on the ``lowest energy"  floor does not necessarily imply  the $\mbox{su}(4)$ singlet on the ``highest
energy" floor,  and \textit{vice versa}.

\section{Examples\label{example}}
In this section we  provide examples of typical, atypical and the BPS representations  discussed in the previous sections. We use two different notations
for $\mbox{su}(2)\oplus\mbox{su}(4)$ in this section. For a few simple representations we just give the dimensions for $\su(2)\oplus\su(4)$, like
$(d,d^{\prime})$. But as we come to study larger multiplets it is better to use the Dynkin labels. Instead of using the full  Dynkin labels  we mostly
give the Dynkin labels of the $\su(2)\oplus\su(4)$  lowest weight only, and the states of different  energies are given in separate lines. Our convention
for $\su(4)$ is that $4 = (1,0,0)$ and the positive fermionic roots $Q$'s are in $(2,4)$. For larger multiplets of $\su(4)$ we give here a short
dictionary
\begin{equation}
\begin{array}{c}
4=(1,0,0)\,,~~~~~\overline{4}=(0,0,1)\,,\\
{}\\
6=(0,1,0)\,,\\
{}\\
10=(2,0,0)\,,~~~~~\overline{10}=(0,0,2)  \\
{}\\
15=(1,0,1)\,, \\
{}\\
20=(1,1,0)\,,~~~~~20^{\prime}=(0,2,0)\,,~~~~~\overline{20}=(0,1,1)\,,\\
{}\\
36=(2,0,1)\,,~~~~~\overline{36}=(1,0,2)\,.
\end{array}
\end{equation}
\subsection{Typical and atypical representations from $\su(2)\oplus\su(4)$ singlet}
First we consider  the supermultiplets where  the superlowest weight is a $\su(2)\oplus \su(4)$ singlet.  The construction is especially simple in  this
case since there are no possible hidden $\su(2)\oplus\su(4)$ lowest weights and  all the possible lowest weights are those listed in
(\ref{lowestweights}).  For generic values of $E_0\neq  0, \pm \frac{1}{3}\mu, -\frac{2}{3} \mu, - \mu$, the representation becomes typical having the
minimal dimension, $256$. In this case the  supermultiplets are of the form  presented in Table 1.

\begin{table}[htb]
\begin{center}
\begin{tabular}{cccc}
\label{single} Floors &
\\
\hline
8 & $(1,1)$    \\
7 & $(2,\overline{4}) $  \\
6 & $(3,6) \oplus (1,\overline{10}) $ \\
5 & $  (4,4)\oplus(2,\overline{20}) $ \\
4 & $(5,1) \oplus (3,15) \oplus (1,20^{\prime})$ \\
3 & $(4,\overline{4})\oplus (2,20)$  \\
2 & $(3,6) \oplus (1,10)$  \\
1 & $(2,4)$   \\
0 & $(1,1)$    \\
\hline
\end{tabular}
\end{center}
\caption{An example of typical representation constructed from $(1,1)$} \label{T1}
\end{table}
The general  typical representation can be obtained from (\ref{finite}) as  the direct product of the minimal typical  supermultiplet  above with an
arbitrary $\su(2)\oplus\su(4)$ multiplet.  Surely the $\su(2)\oplus\su(4)$ multiplet becomes the zeroth as well as the eighth floors in the generic
typical supermultiplet.

Now we consider the atypical representations where the  superlowest weight is a $\su(2)\oplus \su(4)$ singlet.  Using the atypicality condition
(\ref{atypical}), it is easy to see that the representation becomes short when $E_{0} = 0, \pm \frac{1}{3}\mu, -\frac{2}{3} \mu, - \mu$.   For the
unitary representations we choose the sign of $\mu$ such that $E_0$ is nonnegative. All together there are five short unitary  representations which have
$(1,1)$ as the lowest or the highest ``energy" floor. They are presented in Table 2.

\begin{table}[htb]
\begin{center}
\begin{tabular}{cccccc}
Energy$\times(12/|\mu|)$ &&&&& \\
\hline
${12}$ &&&&& $(1,1)$ \\
${11}$ &&&&& $(2,4)$ \\
$10$ &&&&& $(3,6)\oplus(1,10)$ \\
$9$&&&&& $(4,\overline{4})\oplus(2,20)$ \\
$8$ &&&$~(5,1)~$&$(1,1)$&$~(3,15)\oplus(1,20^{\prime})~$ \\
$7$ &&&$(4,\overline{4})$&$(2,4)$&$(2,\overline{20})$ \\
$6$ &&&$(3,6)$&$~(3,6)\oplus(1,10)~$&$(1,\overline{10})$ \\
$5$ &&&$(2,4)$&$(2,20)$& \\
$4$ &&$(1,1)$&$(1,1)$&$(1,20^{\prime})$& \\
$3$ &&$(2,4)$&&& \\
$2$ &&$~(1,10)~$&&& \\
$1$ &&&&& \\
$0$ & $~(1,1)~$ &&&& \\
\hline
\end{tabular}
\caption{Atypical unitary representations from $(1,1)$}
\end{center}
\label{T2}
\end{table}

\subsection{Adjoint representation}
Here we consider the adjoint representation of the superalgebra which is also a kind of atypical representations. We already know the contents of the
adjoint representation: It has 19 bosonic and 16 fermionic generators, and all the bosonic generators commute with the Hamiltonian while the 16 fermionic
generators have eigenvalues, $\pm \ft \mu {12}$.

The zeroth floor is  $(2,\overline{4})$ with the superlowest weight, $|\Lambda_{L}\rangle=\bar{Q}^{24}$ of the Dynkin label, $(1;0,0,1)$. Now acting $Q$,
which is in $(2,4)$, the first floor  decomposes into \be (2,4) \otimes (2,\overline{4}) = (1\oplus 3,\,1\oplus 15)\,. \ee Among them $(3,15)$ decouples
since the lowest weight vanishes, $T^1_1 \gf=0$ because $c_1 = 0$ with $E_0 = -\ft \mu {12}$, $J_z=1,\,J_{u}=J_{v}=0,\,J_{w}=1$ implying the atypicality.
Other three hidden lowest weights are constructed from the superlowest states as follows,
\begin{equation}
\begin{array}{lcl}
(1,1)  &:& (2Q_{11}\E_{z}-\E_{z}Q_{11})\E_{u+v+w} \gf\,,\\
{}&{}&{}\\
(3,1)  &:& Q_{11}\E_{u+v+w} \gf\,, \\
{}&{}&{}\\
(1,15) &:& Q_{11} \E_z \gf\,.
\end{array}
\end{equation}
Surely they correspond to $\mbox{u}(1)$, $\mbox{so}(3)$ and $\mbox{so}(6)$ generators respectively.

Similarly from Table 1 the second floor decomposes  into
\begin{equation}
\begin{array}{cll}
(3,6) \otimes (2,\overline{4}) &=& (2\oplus 4, \,4 \oplus \overline{20})\,,\\
{}&{}&{}\\
(1,10) \otimes (2,\overline{4}) &=& (2,\, 4 \oplus 36 )\,,
\end{array}
\end{equation}
and the possible lowest weights  are
\begin{equation}
\begin{array}{cll}
(2,4) &:& Q_{12} Q_{11} \E_z \E_{v+w} \gf\,, \\
{}&{}&{}\\
(2,\overline{20}) &:& Q_{12} Q_{11} \E_z \gf\,,  \\
{}&{}&{}\\
(4,4) &:& Q_{12} Q_{11} \E_{v+w} \gf\,, \\
{}&{}&{}\\
(4,\overline{20}) &:& Q_{12} Q_{11} \gf\,,\\
{}&{}&{}\\
(2,4) &:& Q_{21} Q_{11} \E_{u+v+w} \gf\,, \\
{}&{}&{}\\
(2,36) &:& Q_{21} Q_{11} \gf\,.
\end{array}
\end{equation}
As before one should check whether they decouple or not.  It turns out that all of them decouple except  only one lowest weight, namely $Q_{21} Q_{14}
\gf=Q_{21} Q_{11} \E_{u+v+w} \gf=-\frac{\mu}{3}Q_{11}$, as it should be. The complete adjoint representation is presented in Table 3.

\begin{table}[htb]
\label{adj}
\begin{center}
\begin{tabular}{cc}
Energy &   \\
\hline
$+\mu /{12}$ & $(2,4)$ \\
0 & $(1,1)\oplus(3,1)\oplus(1,15)$ \\
$-\mu /{12}$ &$(2,\overline{4})$ \\
\hline
\end{tabular}
\caption{Adjoint representation of $\mbox{A}(1|3)$}
\end{center}
\label{T3}
\end{table}

\newpage
\subsection{BPS multiplets}
In this subsection we construct several examples of the  BPS multiplets. In addition to Vacua which have the trivial representation given by a single
state,  there are $3/4, 1/2, 1/4$ BPS multiplets.

We first consider the $\mbox{SU}(2)$ singlet BPS multiplets.  For simplicity we consider the cases where the superlowest weights have only one nonzero
entry for the $\su(4)$ Dynkin labels, $(J_{u},J_{v},J_{w})$. The method is essentially the same as the one employed previously for the construction of the
adjoint representation. We just report the results here.

For 3/4 BPS multiplets we already showed that  only $Q_{14},Q_{24}$ act nontrivially on the superlowest weight.  So the nonvanishing lowest weights on
higher floors are only $Q_{14} \gf, Q_{24}Q_{14} \gf$. The complete 3/4 BPS multiplets are presented in Table 4.

\begin{table}[htb]
\begin{center}
\begin{tabular}{cc}
Energy$\times(12/|\mu|)$ &  \\
\hline
$J+2$   & $(0;0,0,J-2)$ \\
$J+1$  & $(1;0,0,J-1)$ \\
$J$  & $(0;0,0,J)$ \\
\hline
\end{tabular}
\caption{3/4 BPS multiplets}
\end{center}
\label{T4}
\end{table}

Note that when $J_w=1$ the second floor vanishes and the BPS multiplet becomes the fundamental representation.  With the notation denoting the
$\su(2),\su(4)$ dimensions, we present the fundamental representation in Table 5.

\begin{table}[htb]
\begin{center}
\begin{tabular}{lc}
Energy &  \\
\hline
~~$|\mu|/{6}$  & $(2,1)$ \\
~~$|\mu|/{12}$  & $(1,\overline{4})$ \\
\hline
\end{tabular}
\caption{Fundamental representation of $\mbox{A}(1|3)$}
\end{center}
\label{T5}
\end{table}

For  1/2 BPS multiplets, the Dynkin label of the superlowest weights are in general $(0;0,J_v\neq 0,J_w)$. Here for simplicity we treat the cases with
$J_w=0$. The complete 1/2 BPS multiplets are  in Table 6.

\begin{table}[htb]
\begin{center}
\begin{tabular}{cc}
Energy $\times(12/|\mu|)$&   \\
\hline
$2J+4$   & $(0;0,J-2,0)$  \\
$2J+3$   & $(1;0,J-2,1)$  \\
$2J+2$   & $(0;0,J-2,2)\oplus  (2;0,J-1,0)$\\
$2J+1$  & $(1;0,J-1,1)$  \\
$2J$  & $(0;0,J,0)$  \\
\hline
\end{tabular}
\caption{1/2 BPS multiplets}
\end{center}
\label{T6}
\end{table}

\noindent Note that for $J_v=1$ the construction stops at the second floor so we have a shorter multiplet as in Table 7.

\begin{table}[htb]
\begin{center}
\begin{tabular}{cc}
Energy &  \\
\hline
${|\mu|}/{3}$   & $(3,1)$ \\
${|\mu|}/{4}$  & $(2,\overline{4})$ \\
$|\mu|/{6}$  & $(1,6)$ \\
\hline
\end{tabular}
\caption{1/2 BPS multiplet with $J_v=1$}
\end{center}
\label{T7}
\end{table}

Now we consider the 1/4 BPS multiplets. Again for simplicity we start with $(0;J,0,0)$ as the superlowest weight. The complete 1/4 BPS multiplets are
presented in Table 8.

\begin{table}[htb]
\begin{center}
\begin{tabular}{cc}
Energy $\times(12/|\mu|)$& Representations  \\
\hline
$3J+6$  & $(0;J-2,0,0)$  \\
$3J+5$   & $(1;J-2,0,1)$  \\
$3J+4$   & $(0;J-2,0,2) \oplus (2;J-2,1,0)$\\
$3J+3$   & $(1;J-2,1,1) \oplus (3;J-1,0,0)$ \\
$3J+2$   & $(0;J-2,2,0) \oplus (2;J-1,0,1)$\\
$3J+1$  & $(1;J-1,1,0)$  \\
$3J$  & $(0;J,0,0)$ \\
\hline
\end{tabular}
\caption{1/4 BPS multiplets}
\end{center}
\label{T8}
\end{table}

The case of $J_u=2$ reads Table 9.

\begin{table}[htb]
\begin{center}
\begin{tabular}{cc}
Energy &  \\
\hline
${|\mu|}/2$   & $(4,1)$ \\
${5|\mu|}/12$   & $(3,\overline{4})$ \\
${|\mu|}/3$  & $(2,6)$ \\
$|\mu|/4$  & $(1,4)$ \\
\hline
\end{tabular}
\caption{1/4 BPS multiplet with $J_u=1$.}
\end{center}
\label{T9}
\end{table}

\newpage
Finally we consider the $\mbox{SU}(4)$ singlet   BPS multiplets. In addition to the superlowest weight with the Dynkin label, $(J_{z}\neq 0;0,0,0)$, all
the possible $\su(2)\oplus\su(4)$ lowest weights are given by $T_{2}^{l}|\Lambda_{L}\rangle$, $l=1,2,3,4$. The complete $\mbox{SU}(4)$ singlet BPS
multiplets are presented in Table 10.

\begin{table}[htb]
\begin{center}
\begin{tabular}{cc}
Energy$\times(12/|\mu|)$ & Representations  \\
\hline
$2J_z$  & $(J_z;0,0,0)$  \\
$2J_z-1$  & $(J_z-1;1,0,0)$  \\
$2J_z-2$  & $(J_z-2;0,1,0)$  \\
$2J_z-3$  & $(J_z-3;0,0,1)$  \\
$2J_z-4$  & $(J_z-4;0,0,0)$  \\
\hline
\end{tabular}
\caption{$\mbox{SU}(4)$ singlet BPS multiplets}
\end{center}
\label{T10}
\end{table}

From
\begin{equation}
\begin{array}{ll}
(T_{1}^{l})^{\dagger}T_{1}^{l}|\Lambda_{L}\rangle=
\textstyle{\frac{|\mu|}{3}}(J_{z}+1-l)(T_{1}^{l-1})^{\dagger}T_{1}^{l-1}|\Lambda_{L}\rangle\,,~~~&~~l=1,2,3,4\,,\\
\end{array}
\end{equation}
it is obvious that when $J_{z}=1,2,3$ the supermultiplets truncate consistently.


\section{Conclusion\label{conclusion}}
We have identified the superalgebra of the M-theory in the  fully supersymmetric pp-wave background   as  the special unitary Lie superalgebra,
$\mbox{su}(2|4\,;2,0)$ or $\mbox{su}(2|4\,;2,4)$ depending on the sign of the mass parameter, $\mu>0$ or $\mu< 0$ separately. The reason why we have two
inequivalent superalgebras for different signs is essentially due to the fact that in odd dimensions there are two classes of  gamma matrices which are
not related by the similarity transformations. Nevertheless, despite the mathematical distinction, all the physical results we obtained are independent
of the sign.

We have analyzed the root structure of the algebra and presented the second order super-Casimir operator, in the expression of which the $\mbox{su}(2)$
and $\mbox{su}(4)$ Casimirs appear with opposite signs.

We wrote down the general building blocks for the states in any irreducible representation of the superalgebra. The generic supermultiplet  consists of
several floors, from zero to eight at most,  characterized by the discrete energy difference, ${\mu/12}$. Each of the zeroth and the highest floors
corresponds to an irreducible representation of $\mbox{u}(1)\oplus\mbox{su}(2)\oplus\mbox{su}(4)$, while others are in general    reducible
representations. We identified all the ``naturally ordered'' products of the fermionic roots which give the possible lowest weights for the
$\mbox{u}(1)\oplus\mbox{su}(2)\oplus\mbox{su}(4)$ irreducible representations.  Apart from them there are hidden lowest weights depending on the explicit
information of the superlowest weight vector. We have derived explicitly the typicality condition which consists of eight equations in terms of the
energy and other four $\mbox{su}(2)\oplus\mbox{su}(4)$ quantum numbers.

We have completely  classified the BPS multiplets as a special type of the atypical unitary representation. There are two ways of defining the BPS
multiplet compensating the mathematical distinction of the different signs for $\mu$. One is from the superlowest weight and the other from the
superhighest weight.  We find there are $4/16,\,8/16,\,12/16,\,16/16$ BPS multiplets preserving $4,\,8,\,12,\,16$ real supercharges and obtain the
corresponding exact spectrum. We show that, irrespective of the sign of $\mu$, either the ``lowest energy"  floor of the BPS multiplet is  a
$\mbox{su}(2)$ singlet or the ``highest energy" floor of the BPS multiplet is a $\mbox{su}(4)$ singlet.

One can apply the results of the present paper directly to the spectrum of the  matrix model on a pp-wave. The physical states are grouped into the
unitary irreducible representations. In \cite{Kim:2002if,Keshav} it was pointed out that the half BPS multplets with the superlowest weight Dynkin label,
$(0;0,J_{v},0)$ indeed exist in the massive model implying the protected energy from the pertubative corrections to all orders. It is plausible that there
exist more protected BPS or  atypical supermultiplets. Finding such multiplets will be of much interest.

Given the supersymmetry transformation rule for the fermions in the matrix model on a pp-wave, one can in principle obtain the classical BPS equations for
the bosons to satisfy. One of the authors recently developed a systematic method to derive all the BPS equations studying the ``projection matrix" into
the space of the Killing spinors \cite{Bak:2002aq}. In the matrix model for M-theory on a pp-wave the spinors are in the Majorana representation.
Consequently   the dimension of the space of the Killing spinors can be arbitrary from $1$ to $16$ meaning $1/16,\,2/16,\,\cdots, 16/16$ BPS equations.
Our analysis on the BPS multiplets implies that among them only $4/16,\,8/16,\,12/16,\,16/16$ BPS equations admit finite energy solutions. Apart from the
known $1/2,\,1$ BPS solutions \cite{Berenstein:2002jq,Bak:2002rq,Sugiyama:2002jq}, it will be interesting to look for the $4/16,\,12/16$ finite energy
solutions.\newline

\large{\textbf{Acknowledgment}}\\
We wish to thank Gleb Arutyunov, Dongsu Bak, Dongho Moon, Jan Plefka and Piljin Yi for the enlightening  discussion. NK wants to thank KIAS for
hospitality during the visit.


\end{document}